## Abstract

Due to the incentive features of memristors, they are known as a promising candidate for replacing CMOS-based memories, which are faced with various functional challenges in deep sub-micron process technologies. Memristors are non-volatile, low leakage, and dense in comparison to CMOS-based memories, like SRAM. In this regard, resistive RAM (ReRAM) and spin-transfer-torque RAM (STT-RAM) memristors are distinguished among other memristor-based memory technologies due to their superiority in process maturity and metrics such as memory operation energy, memory latency, and area. Hence, this chapter focuses on these two memristor-based memory technologies. Despite the good features of these types of memory, they suffer from some reliability threats. Reliability parameters affect each other, and examining their positive and negative effects has a significant impact on the effectiveness of the proposed solutions. In one view, the threats can be categorized into two classes: 1) read/write error and 2) soft error. In this chapter, we comprehensively describe these threats and present the state-of-the-art solutions that enable the widespread use of memristors, particularly ReRAM and STT-RAM in different applications. Finally, we introduce the emerging ability of memristors as a computing unit aiming to minimize data restoration in computing and show how to perform logic and arithmetic computation in a crossbar array.

# Nanoscale Memristive Devices: Threats and Solutions

Amir M. Hajisadeghi, Javad Talafy, and Hamid R. Zarandi

Department of Computer Engineering, Amirkabir University of Technology (Tehran Polytechnic), Tehran, Iran

## Contents

## 8.1. Introduction

Memristor is the fourth fundamental element in circuit theory which characterizes a nonlinear relationship between magnetic flux and electric charge [1]. The other three circuit elements are resistor, inductor, and capacitor. Although memristor was fabricated in the HP lab in 2008, it has been theoretically presented in 1971 [2, 3]. Memristor has interesting features such as high density, non-volatility, and noise tolerability [4]. Due to the non-volatility of memristors, there would be no need for a power supply to preserve data; thus, it minimizes the leakage power as well. Moreover, the memristor is compatible with CMOS technology so, memristor-based circuits could be implemented in silicon wafers [5].

The primary use of memristors is as memory elements; however, in analog circuits they can be used as programmable resistors as well [6], or new exploitation of memristors is computation [7, 8]. It demonstrates the potential of using memristors in many types of applications [9].

Today a wide range of CMOS-based and memristive memory technologies such as static RAM (SRAM), dynamic RAM (DRAM), phase-change RAM (PCRAM), resistive RAM (ReRAM), spin-transfer-torque RAM (STT-RAM), etc., have been presented. Nevertheless, due to the high leakage, low scalability, and low reliability of CMOS-based memory technologies with transistor feature size downscaling, there is a marked tendency to leverage memristor-based memory technologies [6].

A comprehensive demonstration of popular CMOS and memristor-based memory technologies characteristics have been shown in Table 8.1. In this regard, due to the superiority of ReRAM and STT-RAM over other memory technologies, in terms of area, latency, and energy consumption, the rest of this chapter focuses on these two memory technologies' challenges and solutions as promising memory technologies in the future [10].

Table 8.1. A comprehensive demonstration of memory technologies characteristics.

| ***Memory*** | ***SRAM*** | ***DRAM*** | ***NAND-FLASH*** | ***STT-RAM*** | ***PCRAM*** | ***ReRAM*** | ***FeRAM*** |
|---|---|---|---|---|---|---|---|
| *Non-volatility* | - | - | + | + | + | + | + |
| *Cell area ($F^2$)* | >100 | 6 | 4 (3D) | 6~40 | 4~30 | 4~12 | 15~40 |
| *Supply voltage (v)* | < 1 | < 1 | >10 | < 1.5 | < 3 | < 3 | < 1.8 |
| *Read latency (ns)* | ~1 | ~10 | ~$10^4$ | < 5 | < 10 | < 10 | < 10 |
| *Write latency (ns)* | ~1 | ~10 | $10^5$-$10^6$ | < 10 | ~50 | < 15 | < 10 |
| *Write energy (fJ/bit)* | ~1 | ~10 | ~10 | ~50 | ~$10^4$ | ~50 | 10 |
| *Retention time (year)* | - | ~64*ms* | >10 | >10 | >10 | >10 | >10 |
| *Endurance* | >$10^{15}$ | >$10^{15}$ | >$10^4$ | >$10^{15}$ | $10^8$-$10^9$ | $10^5$-$10^{11}$ | $10^{13}$ |

PCRAM: Phase-change RAM; ReRAM: Resistive RAM; FeRAM: Ferroelectric RAM.

# 8.2. Preliminaries

## 8.2.1. Memristor

Generally, memristor utilizes a relationship between magnetic flux and electric charge. This relationship is defined as the internal state of the memristor by [4],

$$d(\varphi) = Md(q) \tag{8.1}$$

where $M$ is the element's internal state (Memristance), $\varphi$ is magnetic flux, and $q$ is electric charge. Magnetic flux is a function of the electric charge that passes through the memristor. Thus, Eq. 8.1 can be written as,

$$M = \frac{d(\varphi)/dt}{d(q)/dt} = \frac{v}{i} \tag{8.2}$$

where $d(\varphi)/dt$ is the voltage across the memristor, and $d(q)/dt$ is the current. Thus, memristance is the same device as resistance.

The physical structure of the memristor has been shown in Fig. 8.1(a), $w$ specifies the distribution of carriers, and $L$ is the length of the element. In this regard, when applying high voltage to the conducting part, the borderline moves, and the carriers disperse through the length of the element, so the memristance decreases which in the lowest value is known as $R_{on}$. In contrast, the memristance increases by applying reverse voltage, the memristance increases, and the maximum memristance is known as $R_{off}$. Thus, the memristance could be calculated based on the location of the border.

In order to use memristor as a memory cell, low ($R_{on}$) and high ($R_{off}$) resistance values are considered as bit-value "1" ("on" state) and bit-value "0" ("off" state), respectively. Also, memristor-based memory structures are analogous to the conventional crossbar architectures with bit accessing via bit- and word-line, as illustrated in Fig. 8.1(b), which makes memory highly scalable [11].

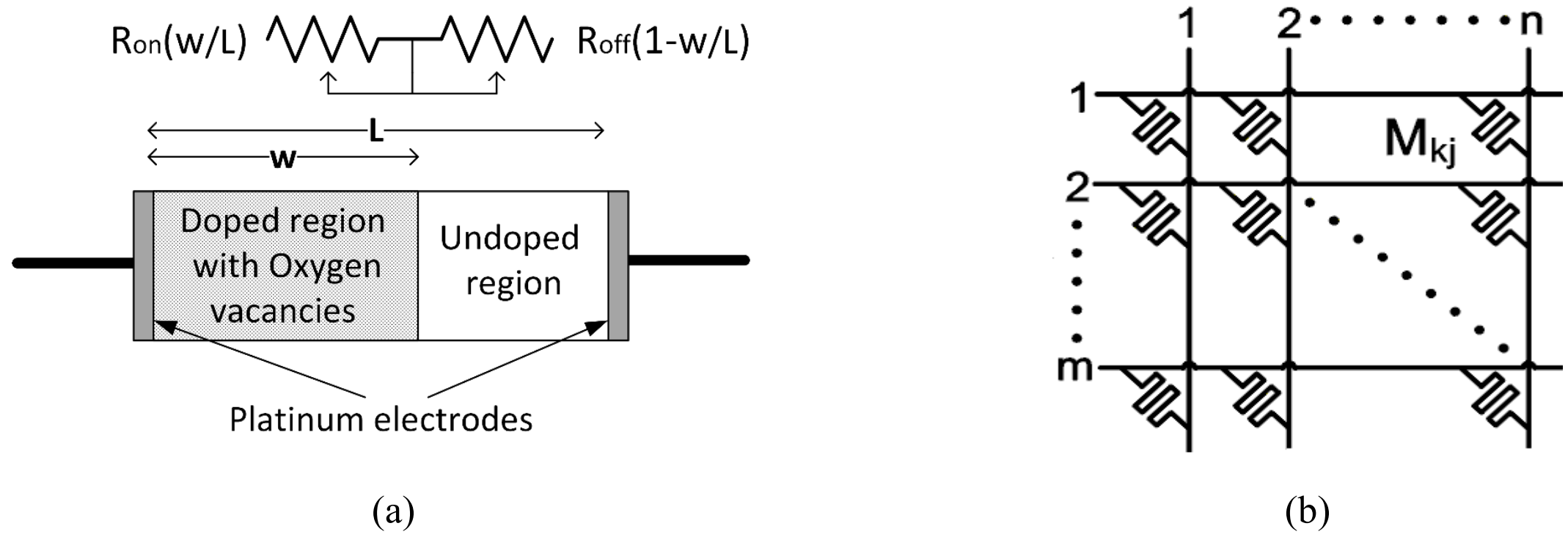

Fig. 8.1. Memristor structure, (a) The physical model of HP memristor, and (b) A crossbar-based memory construction.

### 8.2.2. ReRAM

The resistive random-access memory (ReRAM) cell, as the promising memristive memory technologies, has a simple structure consisting of a metal oxide material sandwiched between two metal electrodes. Depending on the positive or negative voltage applied across the electrodes, the ReRAM cell reaches either a high resistance state (HRS) or low resistance state (LRS), representing logic "0" or "1", respectively.

ReRAM has a better resistance ratio and density than STT-RAM, and its programming current is lower than PCM [12]. In other words, ReRAM is an appropriate candidate in terms of high density and low cost-per-bit due to capabilities like crossbar ReRAM array structure as well as multi-level cell (MLC), which may be the most economical way to have high-density storage. In this regard, it can simply be integrated into the back-end-of-the-line (BEOL) process in which the cell area is as low as $4F^2$ ($F$ is feature size) in stacked form with conventional CMOS or be a multi-level cell that can store multiple bits in a single ReRAM cell [12].

Besides all mentioned advantages for ReRAM (like high density), it is confronted with some deficiencies that need to be solved. In the crossbar array structure, ReRAM is faced with some intrinsic drawbacks named sneak current path and dynamic variation, respectively [13]. Sneak current is the main obstacle to scaling ReRAM arrays and can lead to invalid writing and erroneous data reading. Dynamic variation is another drawback of ReRAM, which is attributed to the stochastic nature of ReRAM, and it is exacerbated in MLC.

Reading a cell in a crossbar ReRAM array is done by applying a $V_{read}$ voltage to the selected word-line and sensing the $V_{out}$ output of a voltage divider circuit from the selected bit line. Then the analog $V_{out}$ voltage will be sensed by an analog to digital converter (ADC) for converting to a binary output. The read margin (RM) as an essential metric affecting the accuracy of the reading is determined by the difference between the highest and the lowest voltages of $V_{out}$ when reading the HRS and LRS cells, respectively, which is defined as Eq. 8.3 [14].

$$RM = \frac{V_{out}(HRS) - V_{out}(LRS)}{V_{read}} \tag{8.3}$$

The multi-level cell capability of ReRAM to store more than one bit in a single ReRAM element has gained increased interest in terms of cost-per-bit and density. However, special consideration should be taken in MLC design due to the stochastic nature of ReRAM. In this regard, many types of research on MLC ReRAM have been conducted in both material-related and system-related areas [15]. For MLC design, enough resistance margin between the intermediate resistance levels and resistance levels uniformity are of great importance. Thanks to the high resistance ratio of ReRAM, the resistance window between the two intermediate levels is large enough to have an MLC design [16].

To address the uniformity problem, some approaches have been presented based on verification-free pulse training techniques favorable in terms of high speed, although they need to variable-multiple voltage sources and intricate control circuitry [17]. In order to cope with the dynamic variation nature of ReRAM, an attached capacitor-based design has been used for both single-level cell (SLC) and MLC designs. This design has benefits such as a few voltage sources, fast response time, and a simple control mechanism [18].

A wide range of experimental investigations in the [19] shows that employing the single/double nanocrack devices in logic design and memory architecture gains superior performance in switching operations. in this paper, the dimensional scaling in the presence of voltage scaling and their correlations are studied. The authors have confirmed that reliable control over switching operations of nanocrack devices is achieved while keeping the operating voltage low.

## 8.2.3. STT-RAM

The fundamental element in STT-RAM memory technology is the MTJ device that stores logic value by its resistance. The MTJ device is a sandwiched tunnel barrier between two ferromagnetic layers. The ferromagnetic orientation in one of the MTJ layers is fix, and the other one is free. Depending on the magnetization orientation of two layers, MTJ state can be either parallel (P) or antiparallel (AP) [20, 21]. Two states of MTJ device structure have been shown in Fig. 8.2(a). Each state of an MTJ device has a low or high resistance called $R_P$ and $R_{AP}$, indicating logic '1' and logic '0'.

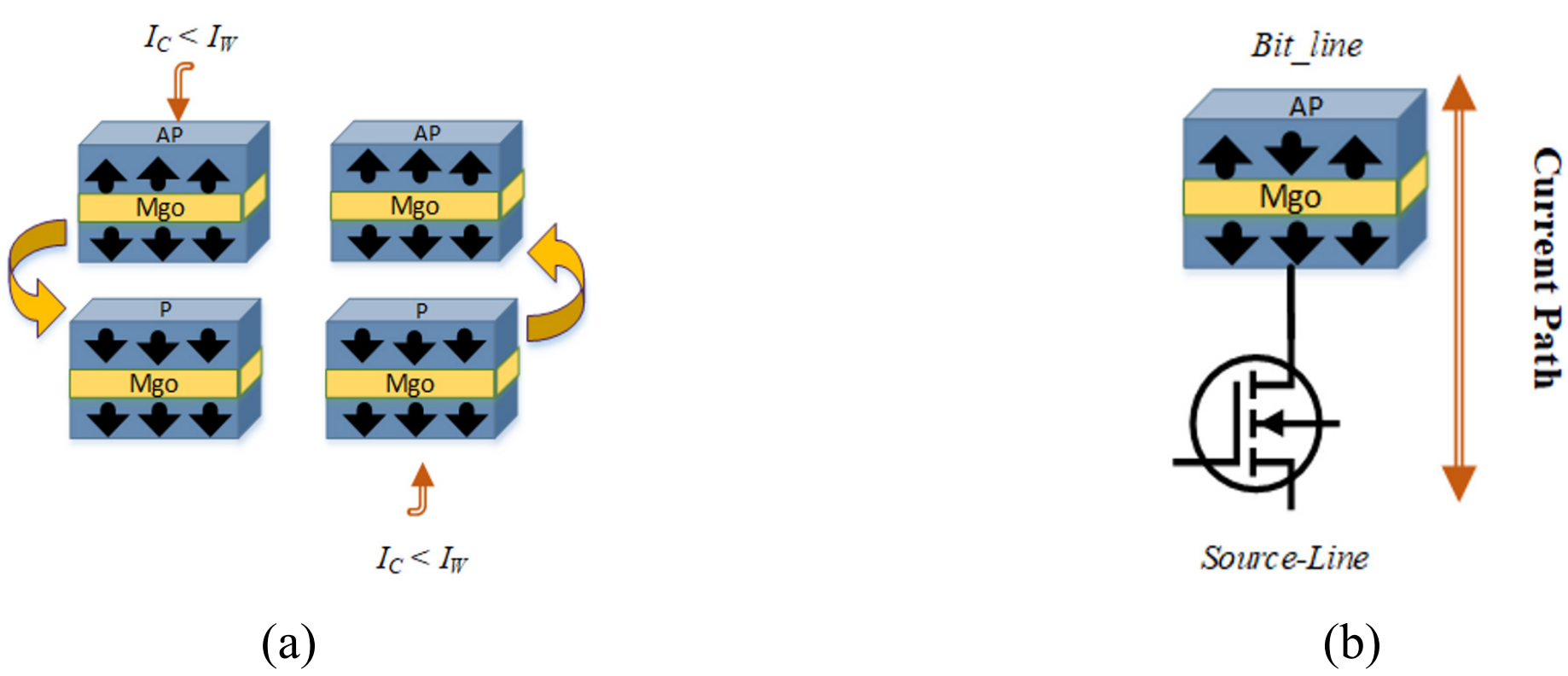

Fig. 8.2. a) MTJ cell and b) basic STT-RAM structure.

The basic structure of the STT-RAM cell consists of a series transistor with MTJ called 1T1MTJ has been shown in Fig. 8.2(b). The word-line of the memory array is connected to the gate terminal of the NMOS access transistor and works like the cell's selector. The read operation in this structure is done by flowing $I_{read}$ (less than $I_C$) from the MTJ and sensed the resistance of the cell; also, the write operation is done by passing $I_{write}$ (greater than $I_C$) from bit-line (BL) to source-line vice versa for writing different logics [20, 22].

## 8.3. Memristor Reliability Challenges and Solutions

With attractive properties which memristor devices incorporate, such as high density, low power, and non-volatility, they are a promising technology for the next generation of VLSI circuits. However, some reliability challenges cause undesired errors, and many efforts are running for the wide-scale industrialization of memristor applications.

Different reliability parameters usually work in conflict with each other, so considering all of them leads to a precise insight for increasing the reliability. For instance, the drawback of ReRAM based crossbar architecture from different views has been investigated in [23] by implying on non-idealities behaviors of nonvolatile devices.

The most important reliability challenges of memristors are categorized into two classes: 1) read/write error, 2) soft error [4]. The following subsections focus on these reliability challenge classes and present state-of-the-art solutions.

## 8.3.1. Memristor Read/Write Error

### 8.3.1.1. Read Error Mitigation

In a crossbar array structure, memristors are positioned on each row-column intersection of the crossbar array, and their resistance (i.e., low or high) represents the stored value. By flowing current through each memristor, the state of that can be sensed. Nonetheless, when current passes through other unwanted cells, causing the memristor to be read erroneously, called the sneak current path problem. The problem leads to reduced read margins and inaccurate detection of cell resistance [24]. To address the sneak current problem, some solutions try to increase the resistance of undesirable paths in parallel with the selected cell aiming to eliminate or decrease the sneak current. Most of the presented solutions aiming to circumvent this problem either eliminate or alleviate the sneak current by using different techniques such as applying a nonlinear selector device beside ReRAM cells. These methods are based on adding a diode or transistor next to the ReRAM cell as a selector [24, 25], shown in Fig. 8.3. However, due to the significant area overhead of aforesaid methods, another class of methods has been presented that modify ReRAM cell structure or read circuits to reduce sneak leakage current [26, 27].

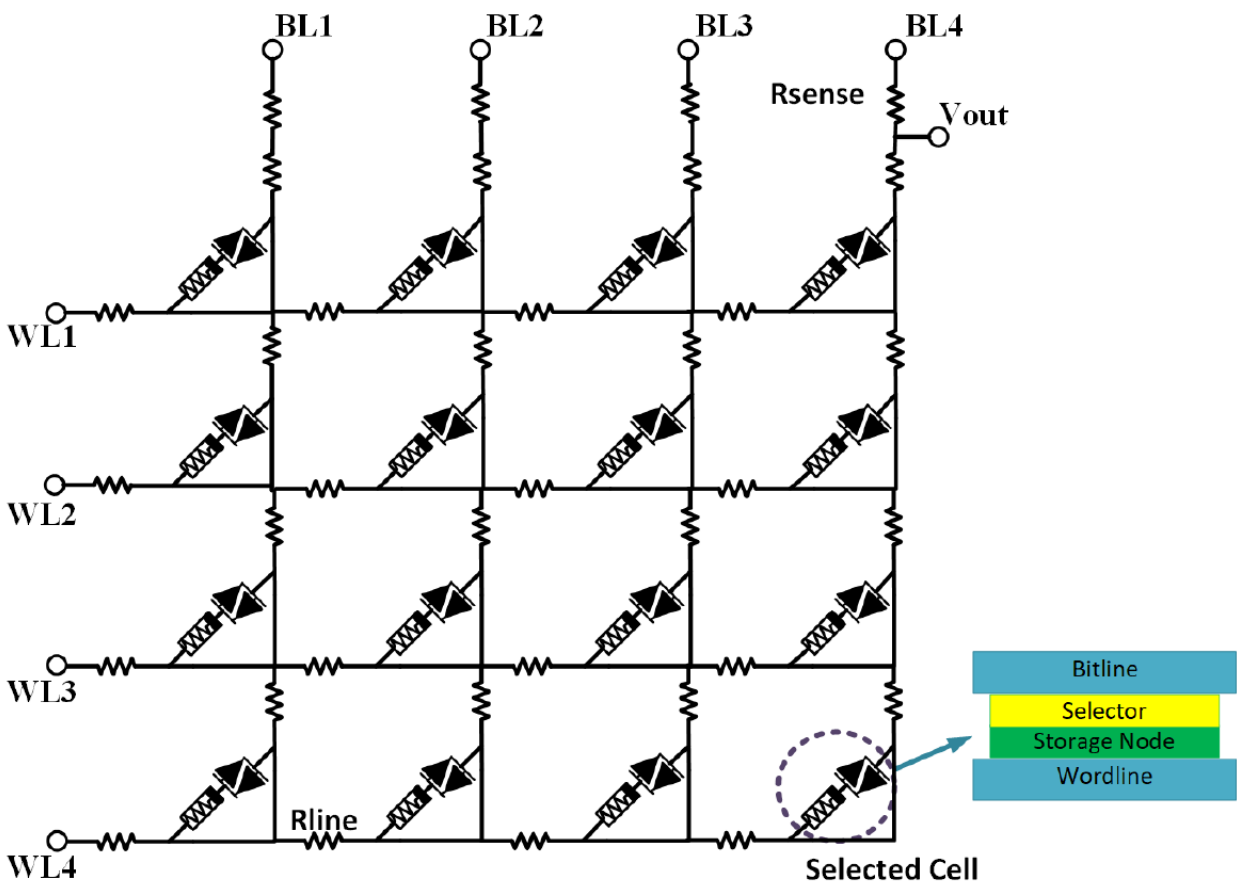

Fig. 8.3. Selector-based crossbar ReRAM array.

In this regard, the proposed method in [28] focuses on changing ReRAM cell structure; the presented cell structure in this research is composed of two resistive elements anti-serially connected. Although this solution could decrease the sneak current, it still suffers from the high area and delay overhead.

In [29], to represent one bit of data, two resistive devices are utilized. Despite this method decreasing leakage current, increasing read margin, and eliminating sensing error, it has a high overhead area. In another work, read and write circuitry has been presented, in which the voltage reference may be adjusted during the read operation, and a sense-before-write approach has been utilized during the write operation [29]. This method has advantages such as improved read margin and reduced sneak current but has disadvantages such as complex read/write circuitry and high latency.

Another work [30] proposed a 3R-2bit cell structure, which has a better area than a complementary ReRAM cell through storing two-bit data in three ReRAM elements. [30] offers a differential 3R-2bit structure, in which each cell is made up of three resistive elements that store two bits of data. This structure inserts HRS resistive elements between two bit-lines to decrease the sneak current. This work improves the read margin up to 46% compared with crossbar architecture, while it is denser, as well. In this regard, 3R-2bit architecture has the same read margin value compared to 2R-1bit architecture with less area per bit. The presented architecture, which gets a good tradeoff among read margin and area overhead, has been demonstrated in Fig. 8.4. As it can be seen, 2-bit data is stored by three ReRAM cells. Moreover, a voltage-based read operation is done to minimize the effect of the sneak current path on the read margin.

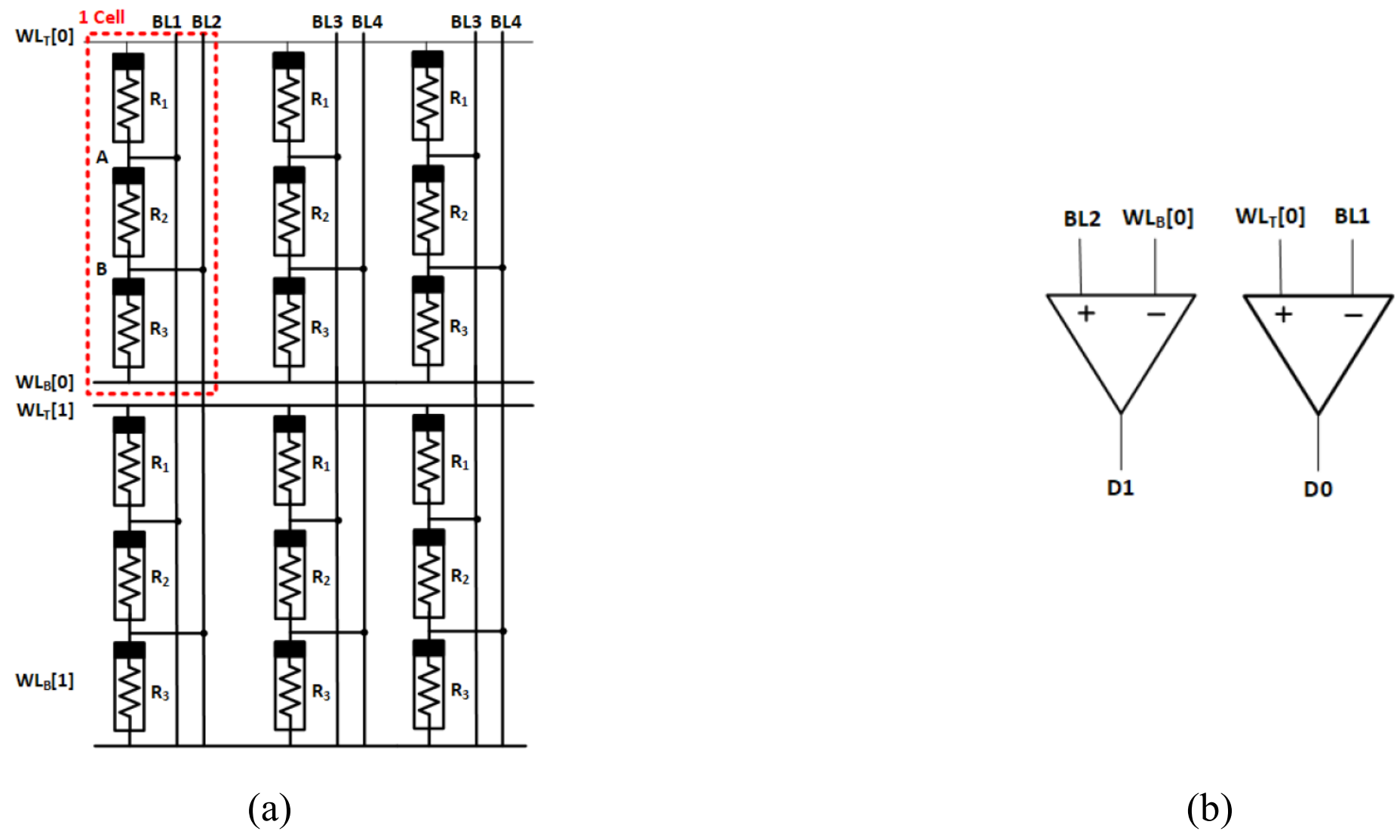

(a) (b)

Fig. 8.4. a) 3R-2bit differential architecture and b) generating of 2-bit data. (*Reprinted with permission from [30].* (*Copyright 2017; IEEE.*))

Based on Fig. 8.4(a), each proposed cell architecture can be accessed by four lines (i.e., $BL_1$, $BL_2$, $WL_T$, and $WL_B$) for read/write operation. To write a value in this architecture, first, reset operation by applying a negative voltage, and second, the set operation has been done by applying a positive voltage to the cell. For instance, to write "00", three resistive elements should be LHL (Low-High-Low resistance) values. Then, $WL_T$, $A$, $B$, and $WL_B$ are connected to $2V_{write}$, $V_{write}$, $V_{write}$, and $GND$, respectively. Table 8.2 demonstrates the write operations conditions for writing all 2-bit values.

Table 8.2. Write operation conditions in 3R-2bit architecture.

| | Write 00 | Write 01 | Write 10 | Write 11 |
|---|---|---|---|---|
| $R_1$ | Set | Set | Reset | Reset |
| $R_2$ | NOP | NOP | NOP | NOP |
| $R_3$ | Set | Reset | Set | Reset |
| $WL_T$ | $2V_{write}$ | $V_{write}$ | 0 | 0 |
| $BL_1$ | $V_{write}$ | 0 | $V_{write}$ | $V_{write}$ |
| $BL_2$ | $V_{write}$ | 0 | $V_{write}$ | $V_{write}$ |
| $WL_B$ | 0 | $V_{write}$ | 0 | $2V_{write}$ |

*(Reprinted with permission from [30]. (Copyright 2017; IEEE.))*

To read data values in this architecture, two $WL_T$ and $WL_B$ are connected to $V_{read}$ and $GND$, respectively. So, the voltage of nodes $A$ and $B$ are calculated by Eq. 8.5, and Eq. 8.6,

$$V_{WL_T} = V_{read} \, , \, V_{WLB} = 0V \tag{8.4}$$

$$V_A = V_{read} \times \left( \frac{R_2 + R_3}{R_1 + R_2 + R_3} \right) \tag{8.5}$$

$$V_B = V_{read} \times \left( \frac{R_3}{R_1 + R_2 + R_3} \right) \tag{8.6}$$

Finally, the 2-bit readout data is generated by comparing the sensed voltage as demonstrated in Fig. 8.4(b).

### 8.3.1.2. Write Error Mitigation

Write error means the cell state does not change to the desired value in a write operation. This issue occurs more often in STT-RAM memory technology versus ReRAM. In this regard, write error occurs due to 1) not enough flowing write current or 2) insufficient time of applying write current to flip the free layer magnetization of MTJ [31]. Generally, write error rate (WER) probability can be quantified by Eq. 8.7 [32],

$$P_{WER}(t) = 1 - \exp\left( \frac{-\pi^2 \Delta \left( \frac{I}{I_C} - 1 \right) \Big/ 4}{\frac{I}{I_C} e^{2\alpha\gamma H_K t \left( \frac{I}{I_C} - 1 \right) \big/ \left(1+\alpha^2\right)} - 1} \right) \tag{8.7}$$

where $I$, $I_{C,}$ and $\alpha$ are, write current value, critical current of MTJ switching, and Gilbert damping factor, respectively. Also, $\gamma$, $\Delta$ and $H_K$ are Gyromagnetic factor, thermal stability factor, and effective anisotropy field, respectively. Based on Eq. 8.7, increasing the $I$ lead to decreasing in WER.

Due to the asymmetric behavior of MTJ, changing the state of MTJ from parallel to antiparallel (PAP) takes less time compared to antiparallel to parallel. Most researches focus on minimizing this asymmetricity and MTJ variation to improve the write error of MTJ based memories like STT-RAM. These researches are classified into two classes: 1) increasing the write current density, 2) increasing the write current time duration. Also, there are two approaches for write completion detection techniques: 1) periodical write tracking (PWT), which works based on write completion checking in fixed time intervals, and 2) real-time write tracking (RWT), in which MTJ state is monitored continuously.

Research [32] is one of the state-of-the-art solutions which works based on the PWT approach. In this research, static and dynamic circuit-level methods have been proposed. The proposed static circuit, shown in Fig. 8.5(a), minimizes the MTJ write asymmetricity. Also, the dynamic method by stepwise increasing the write current tries to minimize WER. The proposed dynamic circuit has been depicted by blue color in Fig. 8.5(b). Although [32] can improve WER, the proposed method faces high power and area overhead due to multiple complex controllers in its dynamic approach.

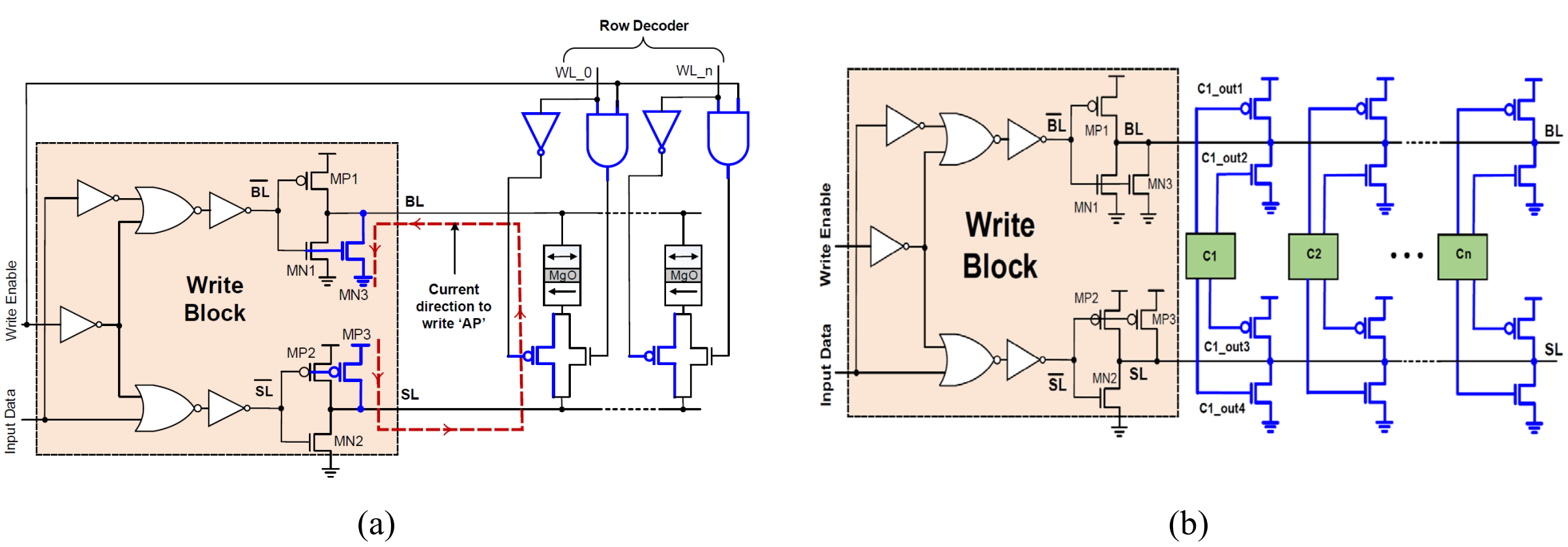

Fig. 8.5. a) Static and b) dynamic write methods presented in [21, 31, 32].

Another research based on the RWT approach has been presented in [33]. In this research, by sensing the *BL* and *SL* voltage changes, the proposed DWT unit stops the write operation as soon as possible after the write completion by enabling implemented transmission gates on *BL* and *SL*. The proposed DWT and memory array structure has been shown in Fig. 8.6.

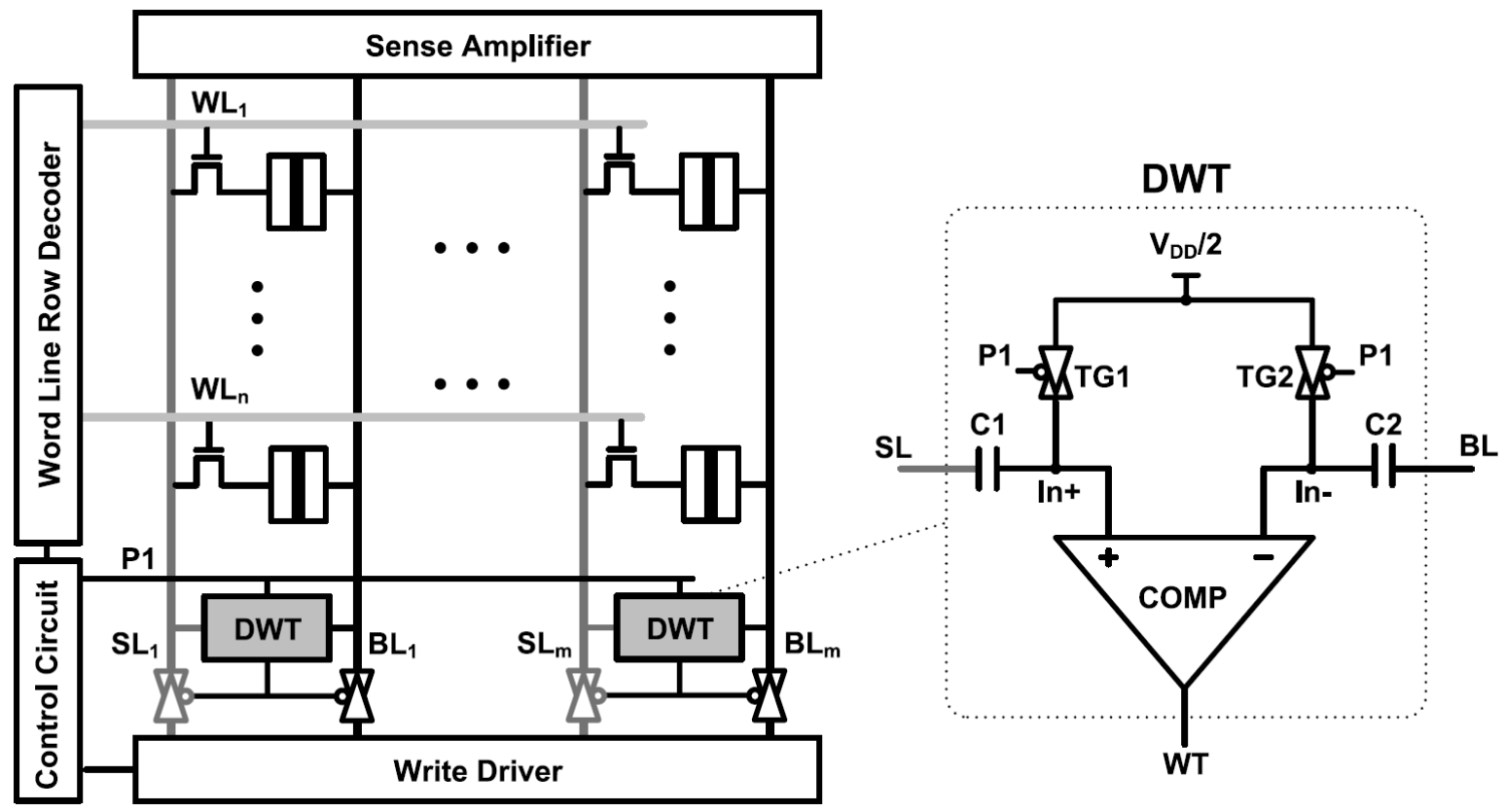

Fig. 8.6. Circuit-level write termination method of [22, 31, 33].

The most significant advantage of this method is its process variation tolerability due to sampling *BL*-*SL* voltage difference before write operation and comparing it with *BL*-*SL* voltages changes while write operation. Nevertheless, high write delay is the drawback of this method.

The method presented in [34] is another work based on the RWT approach to mitigate WER in STT-RAM. It proposes a low power circuit shown in Fig. 8.7 to detect MTJ phase changes (AP→P/P→AP) and eliminate the write current.

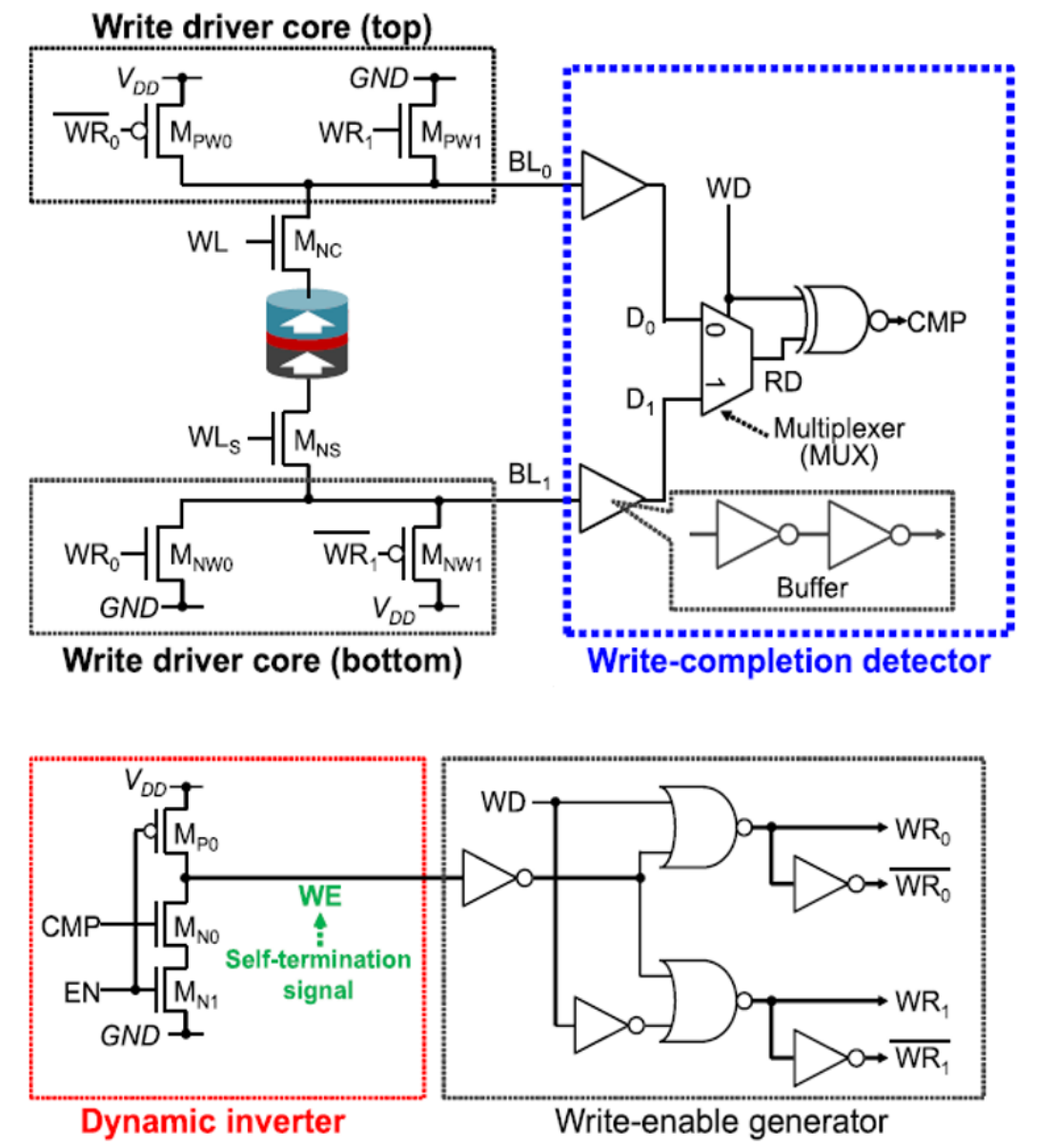

Fig. 8.7. Write and termination circuits presented in [21, 22, 34].

Based on the proposed circuits in Fig. 8.7, if the write data and previous data of STT-RAM cell are the same in the writing process, this condition is immediately detected by the write-completion detector circuit, and related write driver transistors are disabled to eliminate current flow. Also, if write data is different from saved data in memory, the current flows through the memory cell until its state changes, and then the write-completion detector circuit enables the writing process to be terminated. Although [34] presents a fast STT-RAM write circuit, it is worth mentioning that this method's high process variation susceptibility is a critical challenge for the correct functionality of the proposed write-completion detector circuit.

Considering that different current densities are needed for writing logic '0' and '1', in [31], an energy-aware write driver circuit to decrease WER has been proposed. In this regard, a dual-source, continuous-dynamic write circuitry was proposed, which optimizes the physical characteristics of transistors as well. The presented circuitry is divided into two main parts: 1) a thermally-assisted designed circuit for mitigating the asymmetric behavior of STT-RAM to write logics `0' and '1'. 2) A dynamic write current injector designed circuit for speeding up the write operation. To evaluate the results, a comprehensive evaluation compared to the related works has been done. Based on the results, significant improvement in performance and power consumption along with WER reduction has been achieved.

### 8.3.2. Soft Error

Transistors are reaching their scaling limits, and traditional memories face serious design challenges and high leakage current problems. Moreover, by more scaling of transistors, the amount of charge stored in a node decreases due to smaller node capacitance, which leads to more susceptibility to radiation in nanometer circuits. Therefore, the necessity of finding an appropriate alternative is of great importance. As a promising candidate with features like high-bit density, low leakage current, and high radiation tolerance, ReRAM has gained increased attention in recent years. However, deficiencies like susceptibility to radiation strikes in CMOS-based peripheral circuits of the ReRAM arrays need to be maintained by deploying techniques.

### 8.3.2.1. Soft Error Preliminaries

The strike of neutron and alpha particles to a semiconductor device may cause generating electron-hole pairs. These hole pairs in off-state transistors can potentially move in opposite directions in the presence of an electric field leading to the generation of an unwelcome transient current pulse. This transient current can discharge or charge the load capacitance of single/multiple transistors or gates. In this regard, if the collected charge at a load capacitance is more than the critical charge ($Q_{crit}$), it leads to either single event transient (SET) or multiple event transients (METs) [35]. The critical charge is the minimum charge causing a change in the logic value of a circuit's node. When a particle strikes a sensitive node of the storage element directly, the state of the storage element changes depending on the amount of charge deposited, resulting in either a single event upset (SEU) or a single event multi-upset (SEMU) [36]. Unwanted changes due to these phenomena in output results and memory bit-flips are known as soft errors. In this paper, the induced current has been modeled by a double exponential current source defined as Eq. 8.8 [14, 35],

$$I_{inj}(t) = \frac{Q_{inj}}{\tau_1 - \tau_2}\left(e^{\frac{-t}{\tau_1}} - e^{\frac{-t}{\tau_2}}\right) \tag{8.8}$$

where $I_{inj}$ and $Q_{inj}$ are the amounts of the induced current and charge, respectively. Also, $t_1$ and $t_2$ are material-dependent time constants that are selected as 0.16*ns* and 0.05*ns.*

### 8.3.2.2. State-of-the-Art Methods for ReRAM Soft Error Mitigation

Single ReRAM cells are highly radiation-tolerant thanks to their structure. However, they are susceptible to radiation mostly in access transistors and peripheral read/write circuits, which comprise CMOS elements [37]. Most of the circuit-level previous works have been either investigating the radiation effect on ReRAM cells and peripheral circuits in a ReRAM array or seeking solutions to harden the cell [38] and the peripheral circuits. The results revealed that the ReRAM cell is tolerant to total ionizing dose and pulsed-laser irradiation up to hundreds of Krad to several Mrad compared to today's FLASH memories with 75Krad radiation tolerance [37]. However, the drain terminal of the access transistor and peripheral CMOS-based read/write circuits are vulnerable parts. Consequently, the charge generated by energetic particles in radiation-prone areas can result in SEU or SEMU depending on the ReRAM array structure [39].

In this regard, in [40], in order to mitigate soft error, a second transistor is added beside 1T1R ReRAM, and a 2T1R structure is constructed with the cost of the area.

The paper [14] addressed the soft error susceptibility problem of the peripheral read circuits of crossbar ReRAM arrays by applying sizing as a low-cost technique to find either the minimum sizing or one of the best minimum sizing transistors for both SLC and MLC arrays. In this regard, the read circuit has been robust against soft error in two steps utilizing a specific tool written for this purpose. The block diagram of the developed tool's description and sizing methodology has been summarized in Fig. 8.8.

The Monte Carlo simulations have been conducted to analyze the sensitivity of the SLC/MLC array to the soft error. The results revealed that the fully-robust sizing of SLC and MLC would improve soft error susceptibility by a factor of 18.4 and 53.7, respectively. The normalized static power consumption, area per bit, and delay of fully-robust sizing scenarios for SLC and MLC arrays with various sizes from 2×2 to 128×128 have been compared with the default SLC arrays. The results revealed that by using MLC as a high-density structure, the normalized area per bit overhead caused by the sizing of transistors in fully-robust scenarios are diminished, although MLC has static power and 118% delay overhead in comparison with default sizing.

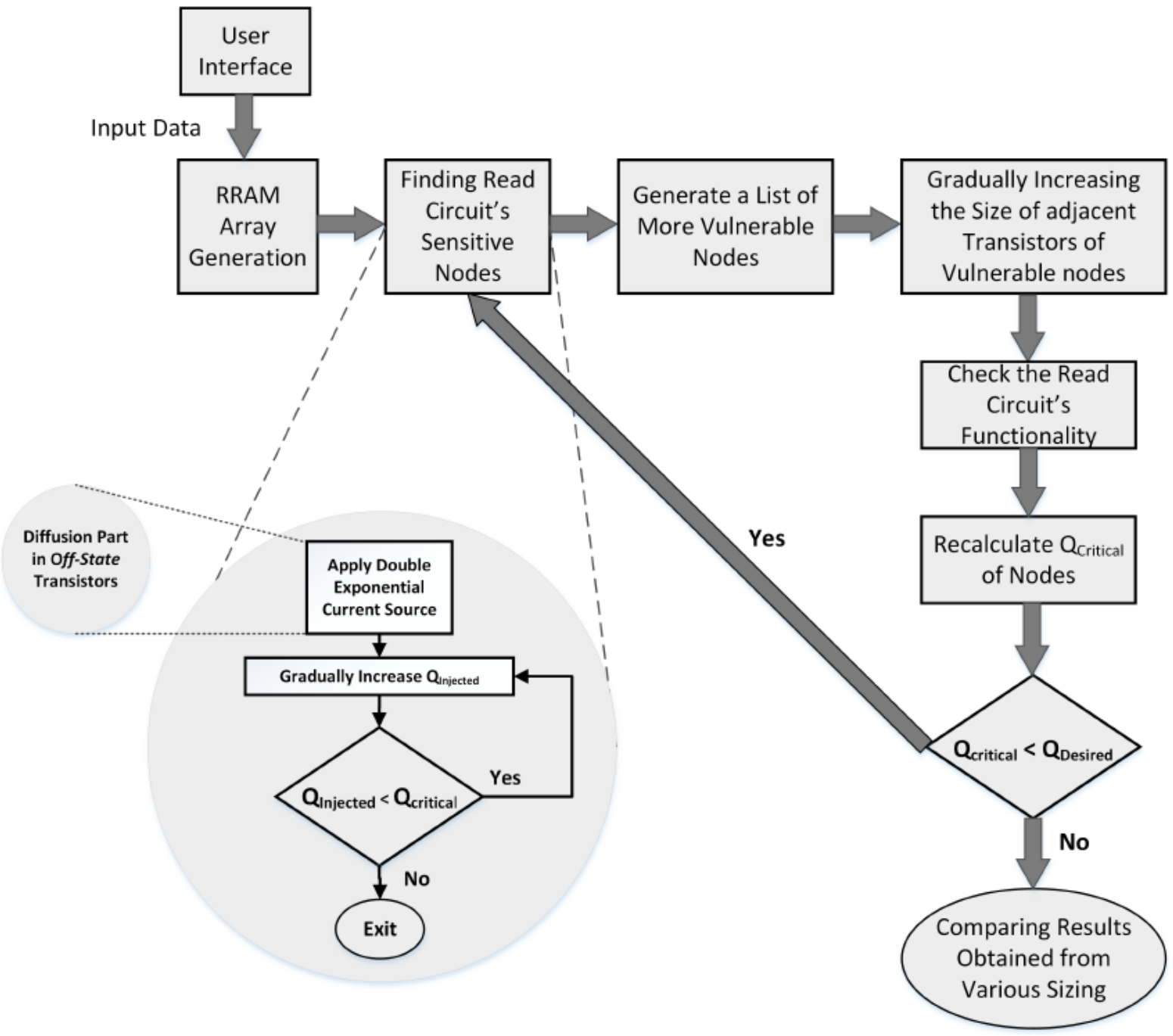

Fig. 8.8. The developed tool's function description and sizing methodology. (*Reprinted with permission from [14]*. (*Copyright 2018; IEEE.*))

### 8.3.2.3. State-of-the-Art Methods for STT-RAM Soft Error Mitigation

In addition to applicability in a memory array structure, STT-RAM can be used in combinational and sequential circuits in which memory elements have been employed to save one or more inputs called Logic-in-Memory (LiM); circuits like Full-Adder, C-element, basic gates (AND, OR, NOT), Flip-Flop, LUT are samples designed based on this structure. It is ideal for building the whole circuit using memory elements; however, this is not yet fulfilled. Thus, LiM is vulnerable to soft error since using the CMOS is not avoidable.

Matsunaga in [41] has investigated LiM-based full adder architecture (Fig. 8.9). The basic form of this architecture includes three parts with different roles: part one is the pre-charge sense amplifier (PCSA) which generates the function output. Part two is made of two MTJs with the ability to rewrite complementary logic. Part three is the controller of the current density which acts as a dynamic current source. PCSA is highly reliable and power-efficient presented in [42]. The read error has been improved by swapping the voltage feedback in part one made of two inverters including $N_{0-1}$ and $P_{0-1}$ instead of using two PMOS transistors $P_{0-1}$ as a simple pull-up. Both PCSA outputs, true and complementary of signals, allow a complete logic family design without extra inverters. This circuit is selected for the investigation of soft error effects on STT-RAM.

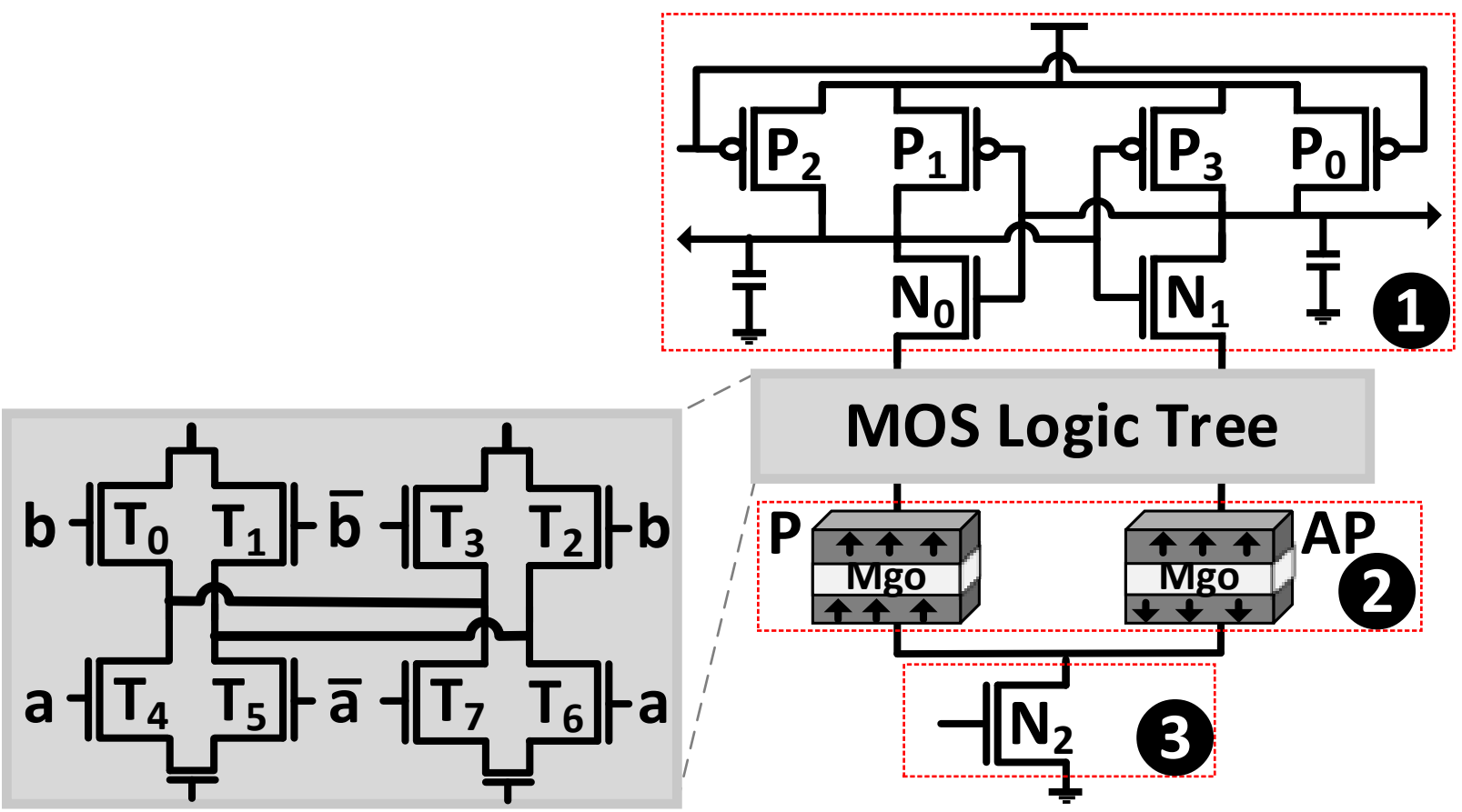

Fig. 8.9. LiM architecture is composed of three parts. (*Reprinted with permission from [38].* (*Copyright 2017; IEEE.*))

In the pre-charge phase, the clock edge is '0', and both outputs reach the VDD value and in the evaluation phase, the clock edge is '1', and the inputs are evaluated. As a result, the full adder circuit (only the summation part) is selected as a test case and investigated in the following scenarios. It should be mentioned that the simulation is base of exponential source as a particle strike (Eq. 8.8) on the circuit node:

A. A particle hits the output nodes of transistors $P_1$ and $N_0$ ($P_3$ and $N_1$).
B. A particle hits the intermediate nodes of $T_0$'s or $T_3$'s drain ($T_1$'s or $T_2$'s drain) depending on being off or on state.
C. A particle that hits the intermediate nodes of $T_4$'s or $T_7$'s drain ($N_5$'s or $N_6$'s drain) depends on being off or on state.
D. A particle hits the intermediate nodes of $N_2$'s drain.

So, to ensure that the circuit is robust, one possible solution is to mask the propagation of error and prevent it from reaching the output [43, 44]. Therefore, a novel circuit is proposed for this purpose (Fig. 8.10).

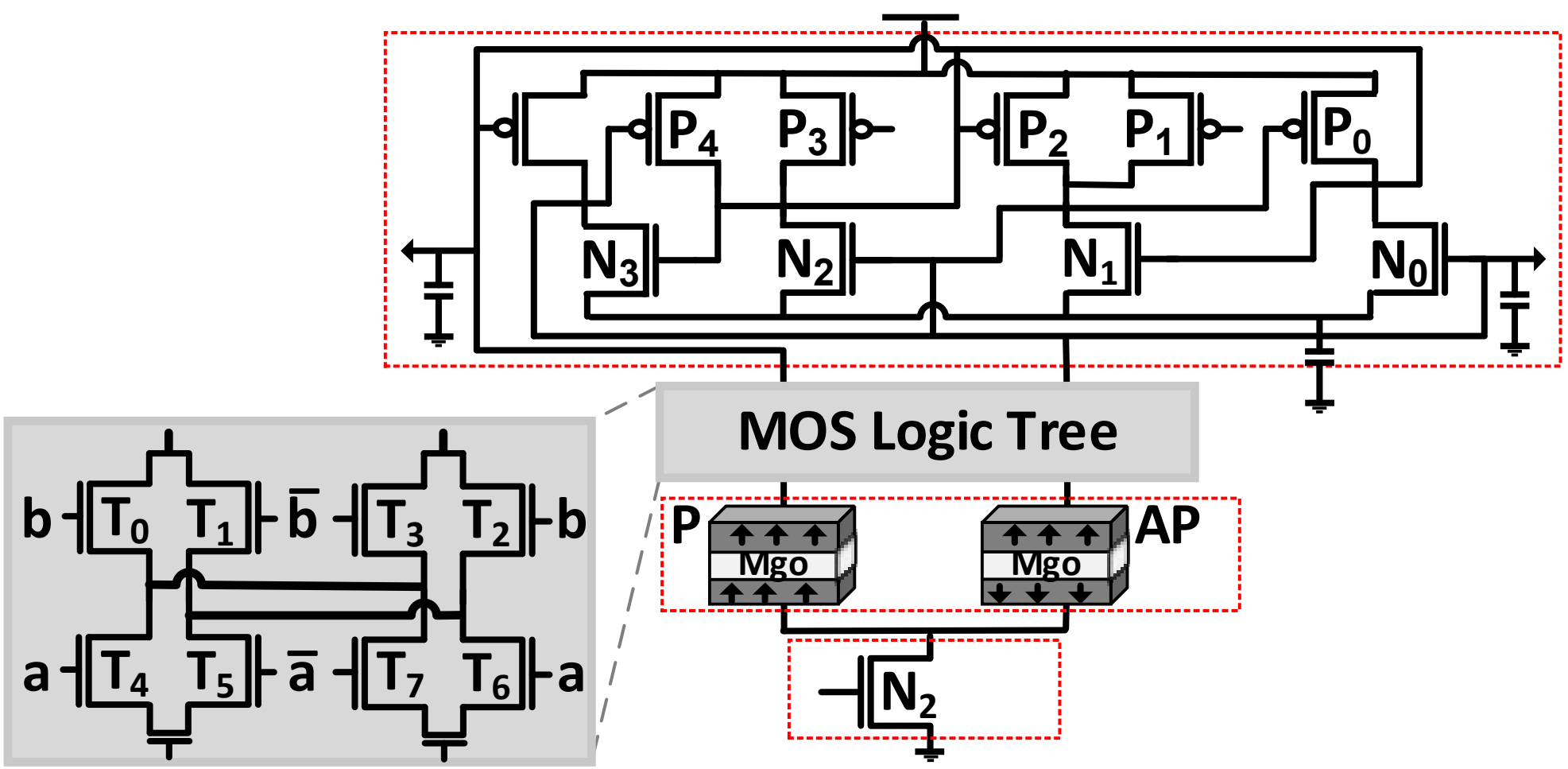

Fig. 8.10. The PCSA is replaced with a robust sense amplifier. (*Reprinted with permission from [38].* (*Copyright 2017; IEEE.*))

As presented in [45], stochastic switching phenomena in MTJ with a created larger domain pulse than the critical threshold could lead to the nonvolatile soft error upset. The proposed circuit reduces the probability of undesirable write because it reduces the value and duration of the pulse (Fig. 8.11). On the other hand, according to smaller MTJ memory device dimensions, the current density required for writing is reducing, which shows that memory in smaller sizes is more impressionable against soft error [46]. The circuit presented in Fig. 8.10 can reduce the pulse width of the SET, which reduces the probability of soft error upset.

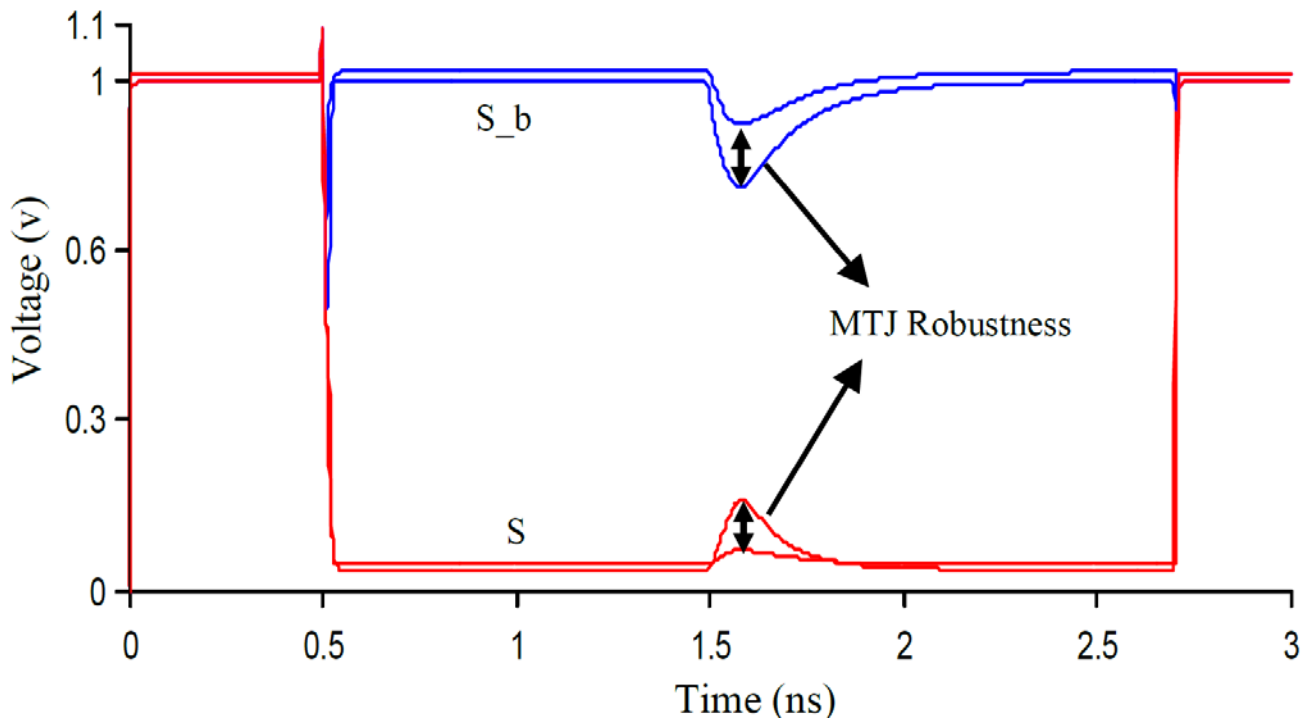

Fig. 8.11. Reduction of the pulse duration of the fault-tolerant circuit. (*Reprinted with permission from [38]. (Copyright 2017; IEEE.*))

Synchronized latch designed based on nonvolatile devices plays an important role, especially in harsh aerospace environments. A novel idea in [47] presents a variation-aware and soft error tolerant latch with 64%, 50%, and 49% improvement in delay, static and total power, respectively.

## 8.4. Memristor Future Direction: Memristor as Logic

The LiM architecture can bring inputs and functions closer together. Nevertheless, promising memristor features alongside the increasing challenge of CMOS in nanoscale technology with decreasing technology sizes lead to a tendency to use memristors in logic circuit design.

### 8.4.1. Logic-in-Memory Circuits Based on STT-RAM

There are a few numbers of multiple-input and single-output circuits which are designed before. However, circuits with lower inputs might be part of circuits with many inputs that consist of some complex functions. A multi-output function may be a combination of circuits with a single output, which can generate the less costly single functions by sharing some common gates.

Generally, the modular design of circuits concludes to a bigger scheme with multi-bit outputs. Herein, the proposed adder and two conventional circuit design methods, carry ripple adder (CRA), a serial connection of single-output adders, and carry select adder (CSA), are compared to evaluate design parameters. Detailed HSPICE simulations are executed in a way that the transistor model is PTM 16*nm* CMOS and STT-RAM compact model. Simulation results prove the proper functionalities of designed circuits. Additionally, results conform to their advantages, 400% write energy improvement of MTJ device, 250% reduction on average power usage, and halve the path delay and comparable small die area and only about reliability degradation of 4%.

As shown in Fig. 8.12, the *N*-bit adder is constructed in ripple form of *N* nonvolatile full adder modules, which are cascading each stage to the next stage to make out (*N*+1)-bit summation of two N-bit numbers, and each stage sends carry input ($C_{in}$, $C_0$, …, $C_{n-1}$) to the next stage to output the carry signal ($C_{out}$). As shown in Fig. 8.13, the second design approach is the carry prediction adder. At first, the $C_{in}$ is unknown, and it determines the summation ($S_0$, …, $S_n$) and $C_{out}$ results at the end using the multiplexers. For this purpose, a 2-to-1 MTJ-based MUX is designed in [48] (Fig. 8.14), in which *B* is one of the input signals saved in MTJ devices. Two other inputs, labeled as *A*, and *S* are conventional volatile signals and are applied as usual. Input *S* plays the selection role between two N-bit inputs *A* and *B* and passes it to the output stage called OUT. Low power consumption, area, and path delay are benefits of the new MUX compared to the CMOS-based multiplexer. Cascading primary circuits to gain a multi-bit output complex one is not an industrialization method because of intensifying power consumption, area, and path delay.

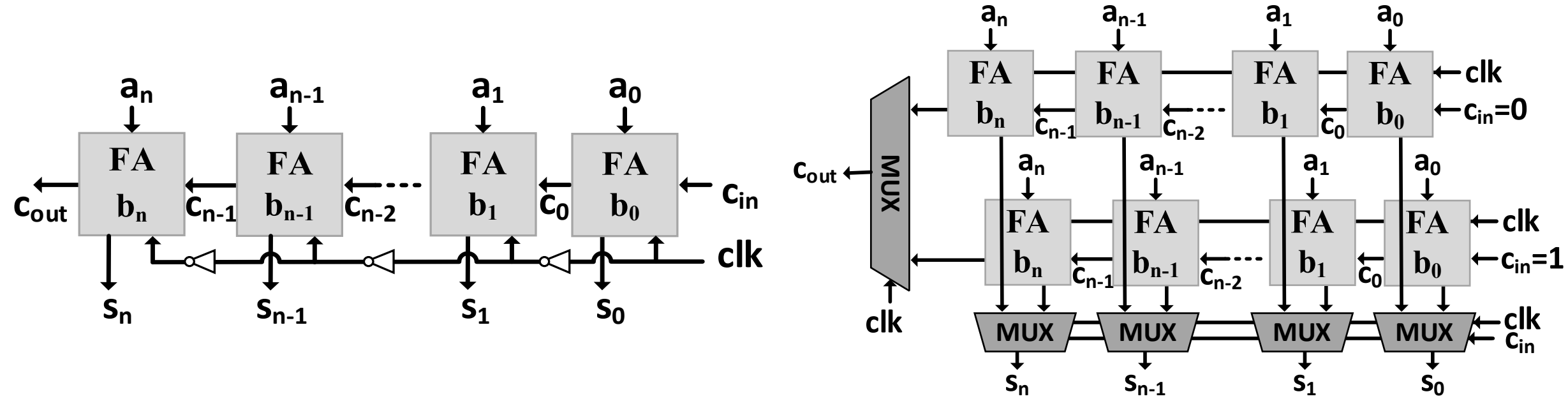

Fig. 8.12. MTJ-based nonvolatile carry ripple adder. (*Reprinted with permission from [48].* (*Copyright 2020; IEEE.*))

Fig. 8.13. MTJ-based nonvolatile carry select adder. (*Reprinted with permission from [48].* (*Copyright 2020; IEEE.*))

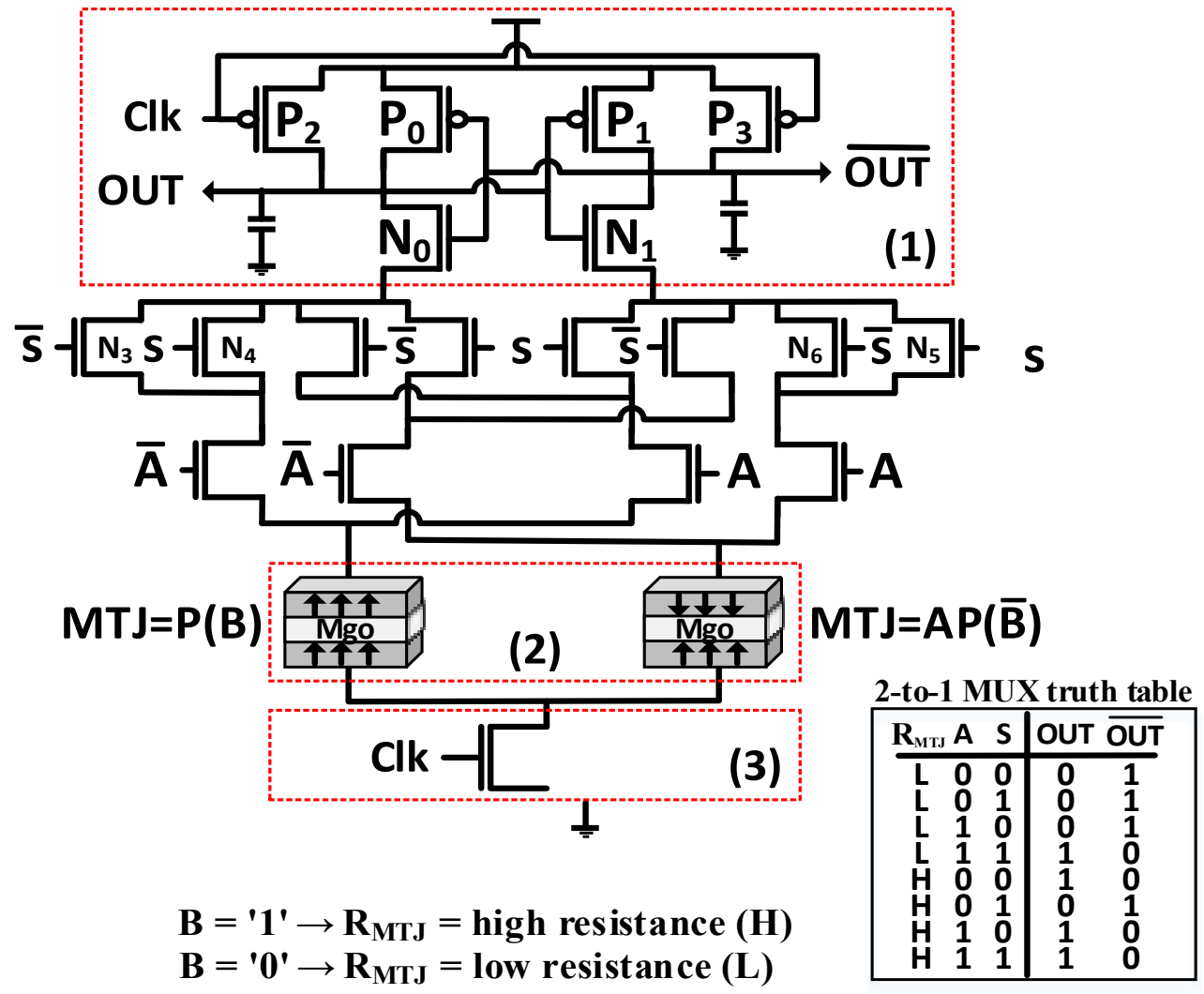

2-to-1 MUX truth table

| $R_{MTJ}$ | A | S | OUT | $\overline{OUT}$ |
|---|---|---|---|---|
| L | 0 | 0 | 0 | 1 |
| L | 0 | 1 | 0 | 1 |
| L | 1 | 0 | 0 | 1 |
| L | 1 | 1 | 1 | 0 |
| H | 0 | 0 | 1 | 0 |
| H | 0 | 1 | 0 | 1 |
| H | 1 | 0 | 1 | 0 |
| H | 1 | 1 | 1 | 0 |

Fig. 8.14. 2-to-1 MUX based on MTJ. (*Reprinted with permission from [48].* (*Copyright 2020; IEEE.*))

### 8.4.1.1. Multi-Output and High-Performance Adder Based on STT-RAM

Two well-known approaches for designing the adder in addition to the advantages and disadvantages of each one was explained in the previous section. Designing an adder in a usual way in which conventional LiM structure is coupled with creating multi-output circuits needs some more modifications to result in an efficient design. The usual way in which rows in Table 8.3 form a track of $n$+1 transistors ($n$ is the input width) serialized to ground the output (connected path) or leave output high. This track includes one stored input in the memory devices (MTJs here) connected to other inputs driving the gate of participant transistors in the track. As shown in Table 8.3, the data inputs of array $A$ ($A_1A_0$) and array $B$ ($B_1B_0$) are applied to the adder coupled with the carry input ($C_{in}$) signal, and the outputs $C_{out}Sum_1Sum_0$ is generated as a final calculation. In the following, sections of the 2-bit adder separately are investigated.

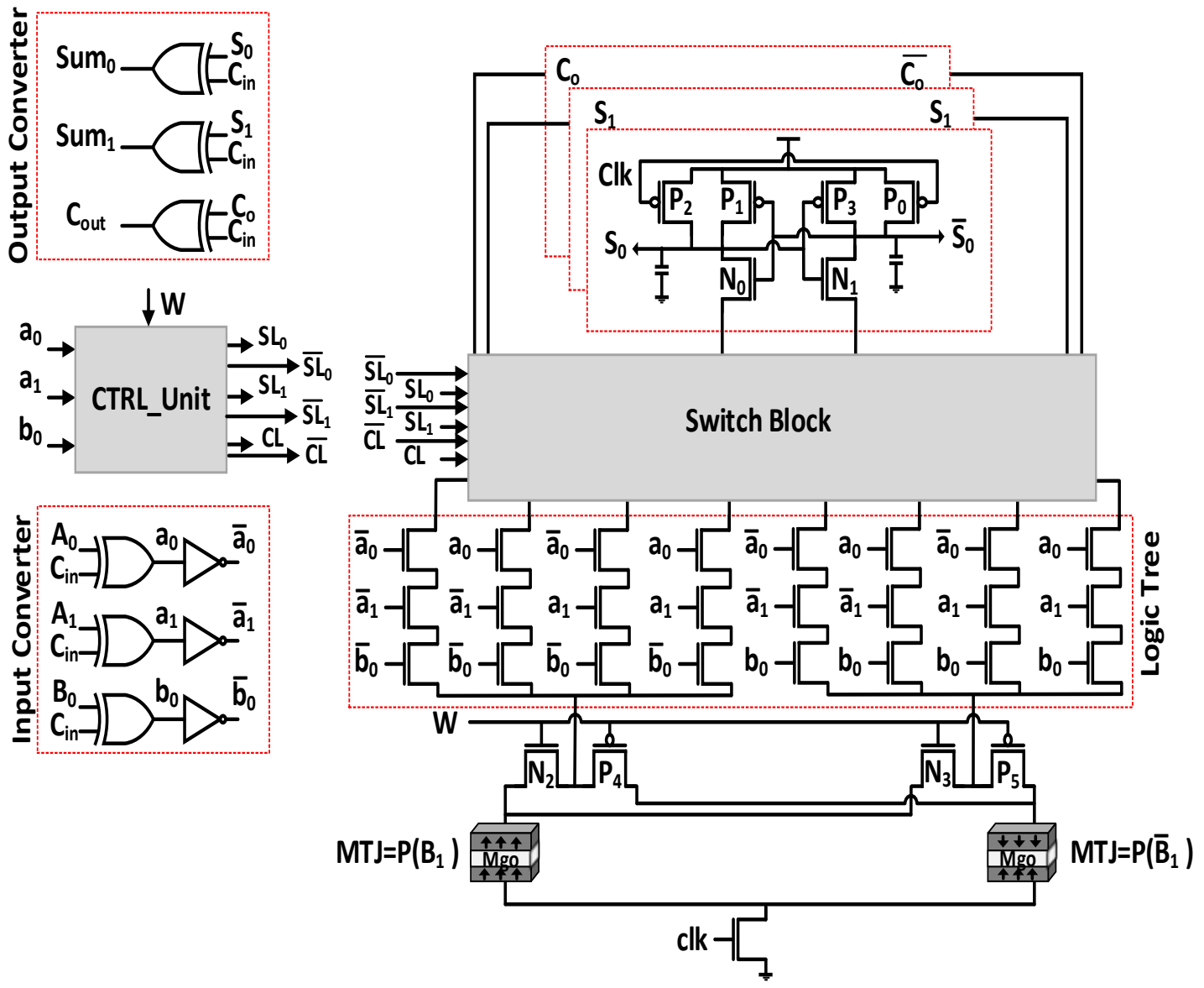

Fig. 8.15. The novel schematic of the designed multi-output 2-bit adder. (*Reprinted with permission from [48].* (*Copyright 2020; IEEE.*))

Table 8.3. Truth table of multi-output 2-bit adder presented in [48].

| Inputs (0 - 15) | Inputs (16 - 31) | Output1 | Output2 |
|---|---|---|---|
| $C_{in}B_1B_0A_1A_0$ | $C_{in}B_1B_0A_1A_0$ | $C_{out}Sum_1Sum_0$ | $C_{out}Sum_1Sum_0$ |
| 0 0 0 0 0 | 1 0 0 0 0 | 0 0 0 | 0 0 1 |
| 0 0 0 0 1 | 1 0 0 0 1 | 0 0 1 | 0 1 0 |
| 0 0 0 1 0 | 1 0 0 1 0 | 0 1 0 | 0 1 1 |
| 0 0 0 1 1 | 1 0 0 1 1 | 0 1 1 | 1 1 0 |
| 0 0 1 0 0 | 1 0 1 0 0 | 0 0 1 | 0 1 0 |
| 0 0 1 0 1 | 1 0 1 0 1 | 0 1 0 | 0 1 1 |
| 0 0 1 1 0 | 1 0 1 1 0 | 0 1 1 | 1 0 0 |
| 0 0 1 1 1 | 1 0 1 1 1 | 1 0 0 | 1 1 0 |
| 0 1 0 0 0 | 1 1 0 0 0 | 0 1 0 | 0 1 1 |
| 0 1 0 0 1 | 1 1 0 0 1 | 0 1 1 | 1 0 0 |
| 0 1 0 1 0 | 1 1 0 1 0 | 1 0 0 | 1 1 0 |
| 0 1 0 1 1 | 1 1 0 1 1 | 1 0 1 | 1 0 1 |
| 0 1 1 0 0 | 1 1 1 0 0 | 0 0 1 | 1 0 0 |
| 0 1 1 0 0 | 1 1 1 0 1 | 1 0 0 | 1 0 1 |
| 0 1 1 1 0 | 1 1 1 1 0 | 1 0 1 | 1 1 0 |
| 0 1 1 1 1 | 1 1 1 1 1 | 1 1 0 | 1 1 1 |

*(Reprinted with permission from [48]. (Copyright 2020; IEEE.))*

In Table 8.3, with a detailed look at outputs of the 8 inputs numbered with $C_{in}B_1B_0A_1A_0$ = 0-7, bits of "$C_{in}B_1$" are constants, the remaining three input bits are counterparts in charging/discharging outputs. The array of "$B_0A_1A_0$" is constructing the logic tree section of Fig. 8.15, in which only one track is active at any time. *CTRL_unit* section is used as a selection of the zero output, which is a participant at grounding. Comparing the inputs $C_{in}B_1B_0A_1A_0$ = 0-7 and inputs ($C_{in}B_1B_0A_1A_0$ = 8-15) in the left column of the table shown in Fig. 8.15 reveals that the only difference is bit $B_1$ particularly stored in MTJs (two red boxes). If inputs of (8-15) are served to the circuit, the outputs 0-8 are retrieved with activation of the MTJs (low resistance) and the track composed in the logic tree section. The switch block section is the connection of desired outputs crossing from a specific track. An approach will be obtaining output from the inputs 8-15 after the input $C_{in}$ is assessed, and if $C_{in}$ is equal to '1', all input bits will be toggled and then will be derived to the circuit. To gain the summation result at the final stage, the output lines from the customized inputs will also be toggled using XOR gates.

Given that the carry bit is '0' in all of the paths for inputs (0-15), there is no need for the carry bit to counterpart in any path. Array $C_{in}B_0A_1A_0$ as Switch Block section inputs are used to select a specified track to determine the output array of the adder circuit. The inputs 0-7 and 8-15 are different in the bit saved in the MTJs. Switch Block outputs are complementary of $CL$, $SL_1$, and $SL_0$. Thus, outputs of the Input Converter section are inputs of this block shown in Fig. 8.15, and signals generated as output of the section *CTRL_Unit* are passed to the inputs $CL$, $SL_1$, and $SL_0$.

### 8.4.1.2. Detailed Design of Synchronous 8-bit Adder Based on STT-RAM

Two related works designed based on spintronic devices (STT-RAM and Racetrack) are an 8-bit adder featured with clock synchronous designed in [49] and an adder designed based on the racetrack in [50]. The idea, designed in [49], presents an adder circuit based on LiM which is fully non-volatile, where all the input are stored in MTJ cells instead of CMOS flip-flops. One of the objections of STT-RAM is the huge write energy and time which is lost over MTJ switching. Overwriting new value in $4\times N$ MTJs ($N$ is input width of Deng adder) degrades energy consumption. The proposed circuit in [49] is not suitable to implement other functions except adder circuits. The work presented in [50] is based on a spintronic memory device called domain-wall (DW) racetrack. It exposes a huge delay of $2.1ns$ only relating to DW motion and nucleation. These two adders can be fabricated with the CMOS fabrication process. For CRA, CSA adders, and the circuit in [49] bit error rate of higher bit orders are more than lower ones; for example, the error rate of $S_1$ is more than $Co$, and the error rate of $Co$ is more than $S_0$, however for the circuit in [48], the bit error rate of all outputs are independent. The value of the next stages in the structures of CRA, CSA, and [49] is not independent of the precedent stage results. For example, if $S_0$ is erroneous, it propagates the error to the next stages; however, $S_1$ and $Co$ are not dependent on each other.

## 8.4.2. Logic-in-Memory Circuits Based on ReRAM

ReRAM crossbar array, in addition to performing memorization operations, can do a set of logic operations. It should be noted that complementary signals in memory applications are non-complementary necessarily. By programming bit-lines and source-line with a proper voltage (*VDD*, *GND*), the path's resistance is configured in a way sense amplifier results the logic correspond to the truth table of logic operation. For example, Fig. 8.16 shows the array and the equivalence of path resistance of ReRAM cells for the XOR gate [51]. Using stateful logic approaches helps the wide industrialization of hybrid CMOS crossbar arrays for retrieving not only arithmetic operations but also making them an ideal target for minimizing the data delay path of storage and process units. Three more works were published in the literature based on the ReRAM [52-54].

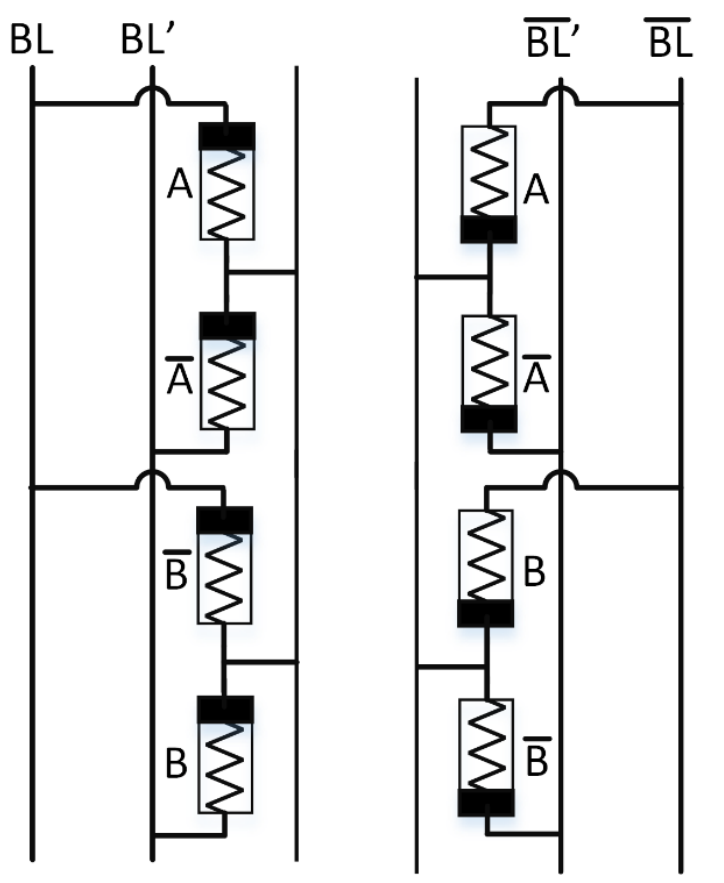

| XOR | | Initial resistance | | states |
|---|---|---|---|---|
| A | B | $R_{BL}$ | $R_{\overline{BL}}$ | |
| 0 | 0 | $\frac{2R_{HRS}R_{LRS}}{R_{HRS}+R_{LRS}}$ | $\frac{R_{HRS}}{2}+\frac{R_{LRS}}{2}$ | $R_{BL} < R_{\overline{BL}}$ |
| 0 | 1 | $\frac{R_{HRS}}{2}+\frac{R_{LRS}}{2}$ | $\frac{2R_{HRS}R_{LRS}}{R_{HRS}+R_{LRS}}$ | $R_{BL} > R_{\overline{BL}}$ |
| 1 | 0 | $\frac{R_{HRS}}{2}+\frac{R_{LRS}}{2}$ | $\frac{2R_{HRS}R_{LRS}}{R_{HRS}+R_{LRS}}$ | $R_{BL} > R_{\overline{BL}}$ |
| 1 | 1 | $\frac{2R_{HRS}R_{LRS}}{R_{HRS}+R_{LRS}}$ | $\frac{R_{HRS}}{2}+\frac{R_{LRS}}{2}$ | $R_{BL} < R_{\overline{BL}}$ |

Fig. 8.16. XOR implementation based on crossbar array [51].

### 8.4.2.1. Multiplexer and Kogge-Stone Adder Using 1S1R ReRAM Crossbar Arrays

Enabling access to the ReRAM devices in the crossbar sequentially, extend the functionality of the memory module from pure memorization to processing unit. The realization of basic Boolean functions, a multiplexer, for example, using 1S1R crossbar arrays, starts from an unknown state *X*. At the first step (a), it should be initialized to zero value. In the next step (b) to store an input with the width of two ($a_1a_0$), two $a_0$ and $a_1$ signals are applied to word-line and '0' to bit-line. In the third step (c), for performing the function of AND with two 2-bit inputs, the first operand is driven to the array, and the second operand with '1' is applied to the word-lines. As shown in Fig. 8.17, the crossbar array is used to implement AND gate logic.

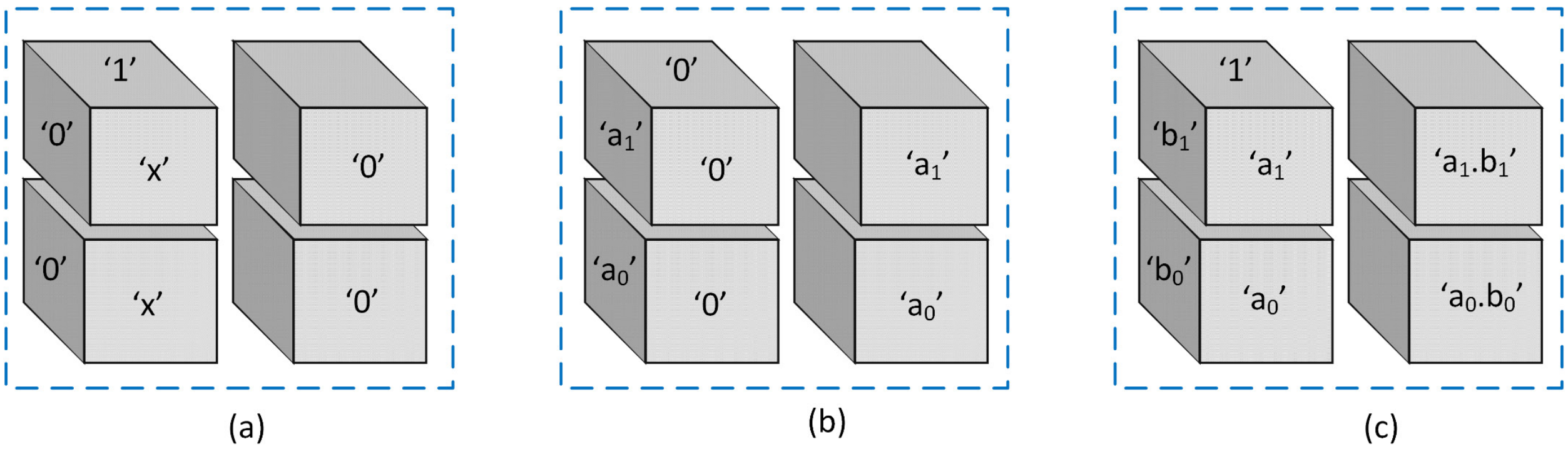

Fig. 8.17. Crossbar array of ReRAM for implementation of AND function in three steps: a) initialization b) load in first input c) load second input [52].

The proposed design is using 2-bit input signals for a 4-to-1 multiplexer. Similarly, a three-input multiplexer is a function including a four-input OR gate and three three-input AND gate. Using the above-mentioned approach designing a multiplexer is simply realized.

Kogge-Stone Adder is another combinational circuit designed based on ReRAM [53]. In the first stage of the adder, the propagate $p$ and generate $g$ terms called the pre-calculation stage, are calculated for each position. The $p$ term shows whether an input carry will be propagated. The $g$ term indicates whether a carry out will be produced. Fig 8.18. shows the architecture of the proposed Kogge-Stone adder.

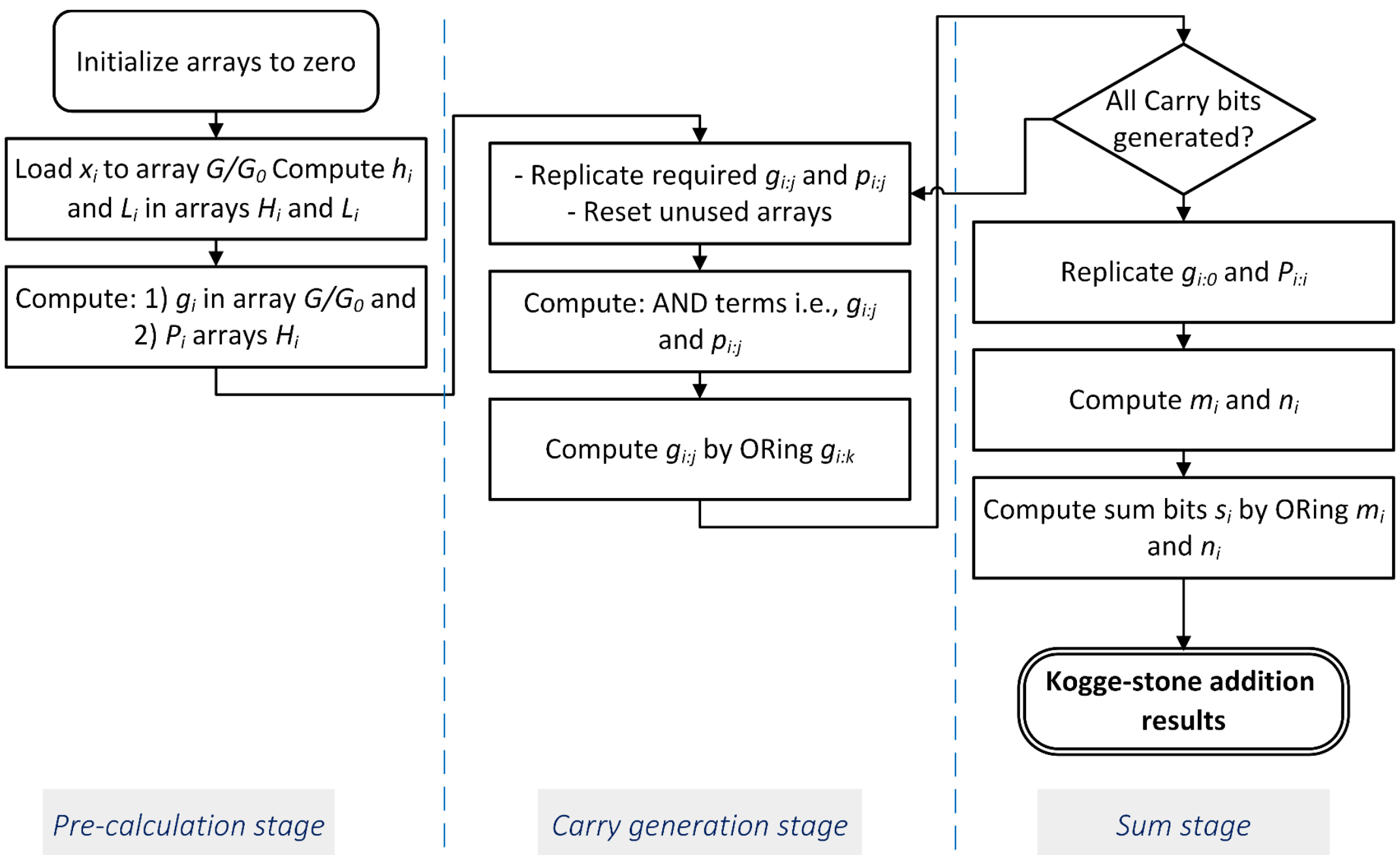

Fig. 8.18. Kogge-Stone adder flowchart for ReRAM-based computation [53].

### 8.4.2.2. Stateful N-bit Full Adder Circuit with Memristive Switches

In this section, a circuit is presented to perform an add operation on a crossbar array by using signal sequences [54]. This circuit consists of 6×(*N*+1) memristor devices, as shown in Fig. 8.19. A common transistor in word-line is required. This transistor raises the flexibility of design to perform different functions as the resistance of the transistor can be tuned. The clock signal is applied to bit-lines which at the same time it encodes the data for function in a serial form. For example, Fig. 8.18 shows a full adder function. Without loss of generality, *A*, *B*, and $C_0$ are the memristive device's data is loaded and represents the usual inputs of the full adder. Before performing the function, *A* and *B* are initialized using a transfer transistor. The carry input is stored in the $C_0$ memristor device. Also, $C_1$ saves the output carry data of the array in the function block, and when the computation is complete, *S* results out the final sum.

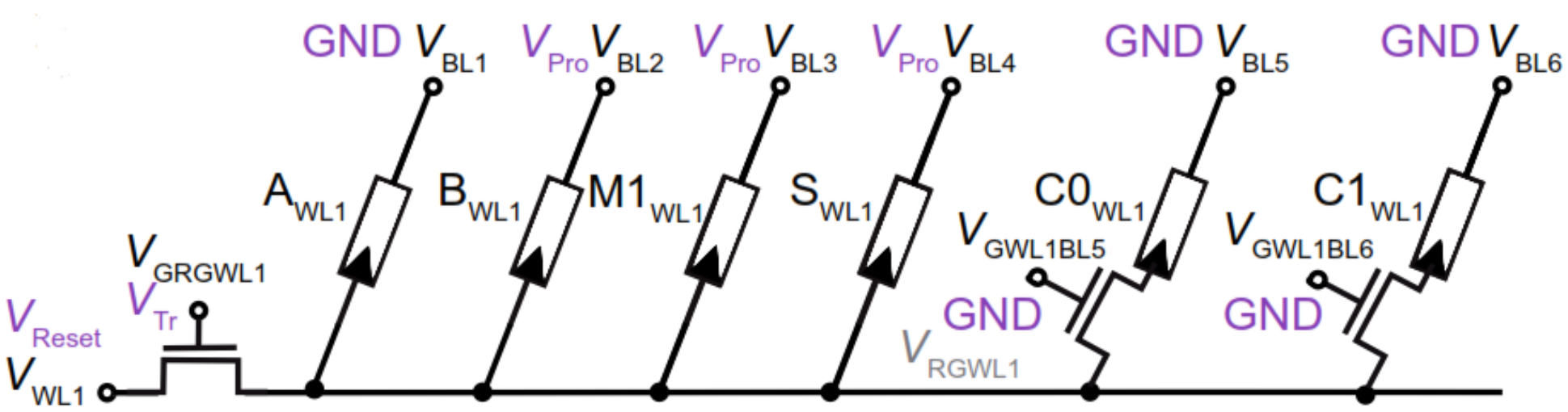

Fig. 8.19. A full adder based on a memristor crossbar array [54].

A full adder was designed as a module with adder functionality; the delay path of input carry to output carry is minimized as much as possible to gain the optimum number of levels needed to perform the adder function. The solution presented here reduces the number of levels by 60%, leading to obtaining high system efficiency.

## 8.5. Summary

In summary, first, we showed the attractive features of the memristors in comparison with CMOS-based memories. Second, the preliminary of ReRAM and STT-RAM as two of the mature memristor technologies, including how they can operate as memory elements were described. Third, we demonstrated the reliability challenges of memristors as a functionality threat. Memristor reliability challenges were classified into two classes of read/write and soft errors. The read/write errors in ReRAM mostly occur due to the sneak current path problem in crossbar array architecture. In this regard, the state-of-the-art solution which addressed this issue by presenting a new 3R-2bit cell structure was described. Also, read/write errors of MTJ-based memories which are originated from the asymmetric and stochastic behavior of MTJs were mitigated by presenting three methods that have superiority over each other in different metrics. Then, the modulation of soft error effects on circuits known as the double exponential current source was introduced and the soft error rate of the ReRAM array was mitigated through a presented circuit hardening methodology based on sizing with negligible overhead. After that, the soft error analysis of the STT-RAM based full-adder circuit was demonstrated. Finally, the logic in-memory paradigm as an emerging feature of memristors which means they can operate as a computation unit along with saving data as memory storage was introduced. An XOR logic, different n-bit adders, and a multiplexer, as well as a state-of-the-art high-performance multi-output adder based on ReRAM and STT-RAM, were presented.

## Index terms